\documentclass[10pt,letterpaper,twocolumn]{article} 

\usepackage{ol2}
\usepackage[draft]{hyperref}
\usepackage{amsmath}

\begin{document}

\twocolumn[ 

\title{Symmetry breaking of optical vortices: Birth and annihilation of singularities in the evanescent field}


\author{P. A. Brand\~ao,$^{*}$ and C. J. S. Juli\~ao}

\address{
Instituto de F\'isica, Universidade Federal de Alagoas, 57.072-970, Macei\'o-AL, Brazil
\\
$^*$Corresponding author: paulo@optma.org
}

\begin{abstract}The interaction of optical vortices (or phase singularities, screw dislocations) with ordinary matter is treated with simple approach. Using total internal reflection phenomenon and superposition of four plane waves incident on a material with refractive index lower than the original propagating medium, we are able to show birth and annihilation of optical vortices in the evanescent field with curved topological features. Until now, this phenomenon has been explored only in free space propagation. By a suitable tuning process, involving the incident angles and the amplitudes of the incident plane waves, it is possible to create unusual topological features of optical vortices in the vicinity of the material. We believe that this work can open new aspects of curved optical vortex manipulation in near-field optics.\end{abstract}

\ocis{260.0260, 260.6970, 050.0050, 050.4865.}

 ] 

\noindent Optical vortices (or phase singularities, screw dislocations) are now well established in the scientific community as basic entities of the electromagnetic field. The first work on optical singularities was carried out by Nye and Berry [1] in the 70's.  Almost twelve years later, singularities were recognized closely related to the orbital angular momentum of a beam of light [2] and now we can say that classical optics is able to describe very well orbital and spin [3] optical angular momentum contributions in a beam of light.

If the phase around a point in the field increases (or decreases) by an integer multiple of $2\pi$, also called topological charge, the phase at the point is undefined. Consequently the field amplitude drops to zero and we have a singularity. The dynamics of topological charges in propagating fields was studied by many authors [4-11]. A useful visualization of the optical vortex is to assign a ``path'' to the point where the phase is undefined, as a function of position. In this way we can construct graphs which show the dark lines representing the points of singularities in the field. For example, the well-know Bessel beam [12] has an on-axis singularity because the phase dependence of the field is given by $\phi (\theta,z) = \beta z + m\theta$ where $\beta$ is the wave vector in the direction of propagation, $\theta$ is the azimuthal angle in cylindrical coordinates and $m$ is the topological charge of the beam. If we plot the function $\phi (\theta,z)$, we observe that a singularity exists (for $m\ne 0$) in the center of the beam ($\rho = 0$) and follows a straight line propagation parallel to the $z$-axis. Also, the surface where the phase is constant (i.e. the wavefront of the beam) is helical. Such helical wavefront give a non-zero contribution to the transversal component of linear momentum. This can be used to move particles around the singularities as opposed to the spin angular momentum which rotates the particles around their own axis [10].

The problem of producing optical beams with vortices paths that are not straight lines, arises naturally and was answered theoretically by M. V. Berry and M. R. Dennis [13] by making interference of Bessel beams and using a perturbative analysis. The theory of Berry and Dennis was verified experimentally by Leach et. al. [14]. In particular, O'Holleran et. al. [15] showed that we can have unusual topological features (loops, births and annihilations) arising with the coherent superposition of three, four and five plane waves. Their work inspired the present paper because a rich variety of singular paths can exist, using only these basic entities of the electromagnetic theory, i.e., plane waves. Non-straight-line propagation of optical vortices is also found in free-propagating Airy beams, which results in parabolic trajectories, as demonstrated by Mazilu et. al. and Dai. et. al. [16,17]

In this work, we used the total internal reflection phenomenon to create evanescent fields with curved topological singularities. The papers cited previously lead us to consider questions like if is it possible to have loops, knots, birth and annihilation of optical vortices in the evanescent field. Also, the question of what  happens to the optical vortices when they interact with matter is studied. Here we show that it is possible to visualize, numerically, birth and annihilation of singularities. The question involving loops and knots remains open to numerical and experimental investigation.

In order to show the relevant aspects of the effects in near-field regime, we take a planar surface separating two media with different index of refraction. Plane waves (four, in this paper) with wave vectors $\vec{k}$ travelling originally in the medium with index of refraction $n_{1}$ making an angle $\theta_{j}$ with the normal, strikes a medium with index of refraction $n_{2}$ giving rise to refracted waves with wave vectors $\vec{k''}$ and reflected waves with wave vectors $\vec{k'}$. If we take $n_{1}>n_{2}$, there exists an angle $\theta_{c} = \sin^{-1}(n_{1}/n_{2})$, namely, the critical angle, such that if the incident plane waves satisfy the condition $\theta_{j}>\theta_{c}$ the cosine of the refracted angles becomes purely imaginary, consequently, the waves remain bounded within the $z$ direction with dependence $exp(-z/d)$, where $d^{-1}=k'(sin^{2}\theta_{j}-sin^{2}\theta_{c})^{1/2}$ is the characteristic length scale. 

The evanescent field in the region $z>0$ can be calculated by [18]
\begin{align}
{E_{ev}(x,y,z>0)} &=\sum_{j=1}^{4}\frac{2n_{1}E_{j0}\cos\theta_{j}}{n_{1}\cos\theta_{j}+(n_{2}^{2}-n_{1}^{2}\sin^{2}\theta_{j})^{1/2}}  \label{Eq1} \nonumber \\
&\quad \times \exp[ik''(x\cos\phi_{j}\sin\theta_{j}')] \nonumber \\
&\quad \times \exp[ik''(y\sin\theta_{j}'\sin\phi_{j}+z\cos\theta_{j}')].
\end{align}
where $\theta_{j}'$  and $\theta_{j}$ are the refracted and incident angle of the $j$-th incident wave, respectively, $E_{j0}$ is the amplitude of the $j$-th incident plane wave and $\phi_{j}$ is the azimuthal angle in spherical coordinates. Using Eq. (1) and varying twelve parameters, the incident angles and the incident amplitudes, we can calculate the evanescent field above the plane for a variety of situations.

Let us take four plane waves propagating from medium with refraction index $n_{1}=1.5$ to the medium with refraction index $n_{2}=1.0$. We assume that the polar angles in spherical coordinates are given by $\theta_{1}=1.2\theta_{c}$, $\theta_{2}=1.5\theta_{c}$, $\theta_{3}=1.6\theta_{c}$, $\theta_{4}=1.7\theta_{c}$, the azimuthal angles are given by $\phi_{1} = 0$, $\phi_{2} = \pi/2$, $\phi_{3} = \pi$, $\phi_{4} = 3\pi/2$  and all four incident plane waves have unit amplitude. In [15] it was shown that if we superimpose four plane waves with equal magnitudes in free space, the singularities are straight lines. So, this is what we get for $z<0$ before the waves strike the surface. But when we look at the field right above the surface, we observe that there are line vortices annihilated, as shown in Fig. 1. The presence of the boundary, which separates two different media, breaks the symmetry of straight line vortex propagation. To explain this result, we should look at the amplitudes of the individual refracted waves. They all depend on $z$, so we have plotted in Fig. 2(a) the dependence of $a_{1}+a_{4}$ and $a_{2}+a_{3}$ ($a_{j}$ is the amplitude of the $j$-th refracted plane wave) as a function of $z$. We clearly see that for any value of $z$ above the surface, the following inequality is satisfied: $a_{1}+a_{4} > a_{2}+a_{3}$. As demonstrated in [15] this condition implies that loops of vortices must be present. The Fresnel's coefficients play the principal role in this discussion because they are the amplitudes of the refracted waves necessary to account for the phasor description (see [15]). Also, we observe that they appear in pairs, as expected for topological charge conservation [19]. Finally, part (b) of Fig. 1 shows the path of the vortices marked with a red and blue circle as a function of $x$, $y$ and $z$. Clearly we see the annihilation effect of vortices appearing solely due to the spatial symmetry  breaking. 

\begin{figure}[htb]
\centerline{\includegraphics[width=5.5cm]{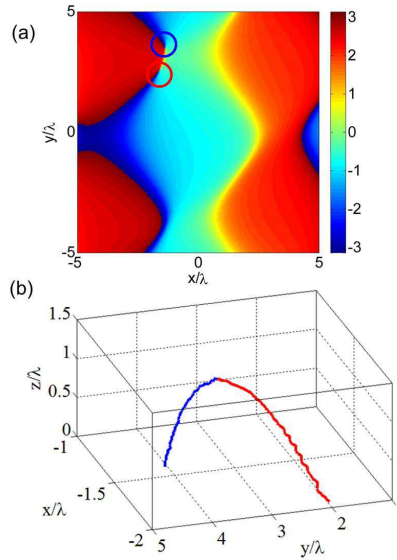}}
\caption{(Color online) Annihilation of optical vortices in the evanescent field. Parameters used in the simulation: $\theta_{1}=1.2\theta_{c}$, $\theta_{2}=1.5\theta_{c}$, $\theta_{3}=1.6\theta_{c}, \theta_{4}=1.7\theta_{c}$, $\phi_{1}=0$, $\phi_{2}=\pi/2$, $\phi_{3}=\pi$, $\phi_{4}=3\pi/2$, $n_{1}=1.5$ and $n_{2}=1.0$. (a) Phase profile of the evanescent field. The blue and red lines indicates singularities with opposed topological charges. (b) The vortex path of the singularities indicated in part (a) as a function of $z$.}
\end{figure}

Creation of pairs of singularities is also possible. In Fig. 3(a) we show the phase profile of the evanescent waves but with different parameters given by $\theta_{1}=\theta_{2}=\theta_{3}=1.2\theta_{c}$, $\theta_{4}=1.7\theta_{c}$, the azimuthal angles are the same as in Fig. 1. and the amplitude of the fourth incident plane wave is taken to be $5.5$. With these parameters the amplitudes of the refracted waves are plotted in part (b) of Fig. 2, so we still have the condition $a_{1}+a_{4} > a_{2}+a_{3}$ satisfied for all $z$ (we were labeling the amplitudes in increasing order,  but in this case there is no difference). But now we can demonstrate the birth effect of singularities with opposed signs, as shown in Fig. 3(b). In the picture of Talbot cell, we are simply moving the cell to contemplate another region which shows the effect. 
\begin{figure}[htb]
\centerline{\includegraphics[width=6.0cm]{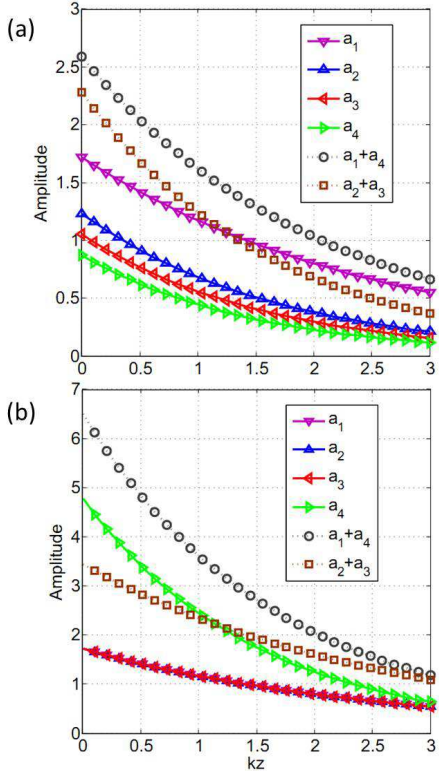}}
\caption{(Color online) (a) Amplitudes of $a_{1}+a_{4}$ and $a_{2}+a_{3}$ in function of $z$ for the annihilation of optical vortices case. (b) Amplitudes of $a_{1}+a_{4}$ and $a_{2}+a_{3}$ in function of $z$ for the birth of optical vortices case.}
\end{figure}
A detailed analysis for the conditions under which we have creation/annihilation of optical vortices is still open to discussion. The difficulty arises because for each $a_{j}$ we have three free parameters ($\theta_{j}, \phi_{j}$ and $E_{j0}$) that, when placed in $a_{1}+a_{4}>a_{2}+a_{3}$, generates an inequality with twelve free parameters (fifteen if we consider the relative initial phases). We are by no means in a position to discuss the general case in the present letter, leaving this interesting analysis for future work. 

Although we have showed that it is possible under suitable tuning process to create and annihilate optical vortices in the evanescent field, the symmetry also breaks even in the condition $\theta_{j}<\theta_{c}$, i.e. for propagating fields beyond the boundary (or for $n_{1}<n_{2}$, in which there is no condition for a bounded wave). We choose to adopt the evanescent-view for its implications in near-field optics, still, to our knowledge, unverified in experimental conditions. Some questions still need to be answered and experiments need to be performed. Also, different materials should lead to different and unexpected curved vortex topological behavior. We do not face these problems in this paper, retaining our results only to show that it is possible to create new degrees of freedom in the optical tweezers area using optical singularities. 
\begin{figure}[htb]
\centerline{\includegraphics[width=5.5cm]{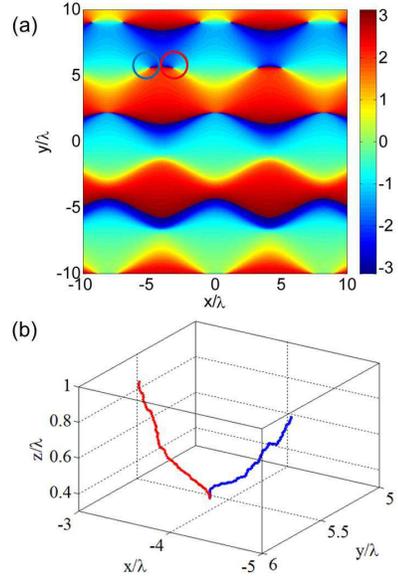}}
\caption{(Color online) Creation of optical vortices in the evanescent field. Parameters used in the simulation: $\theta_{1}=1.2\theta_{c}$, $\theta_{2}=1.2\theta_{c}$, $\theta_{3}=1.2\theta_{c}, \theta_{4}=1.7\theta_{c}$, $\phi_{1}=0$, $\phi_{2}=\pi/2$, $\phi_{3}=\pi$, $\phi_{4}=3\pi/2$, $n_{1}=1.5$, $n_{2}=1.0$ and the amplitude of the fourth plane wave is $5.5$. (a) Phase profile of the evanescent field. The blue and red lines indicates singularities with opposed topological charges. (b) The vortex path of the singularities indicated in part (a) as a function of $z$.}
\end{figure}

In conclusion we showed that evanescent fields can contemplate birth and annihilation of optical singularities by simple tuning process. From a physics standpoint, the symmetry of straight line propagation is broken when they interact with matter, and we showed that the phasor argument is sufficient to characterize the dynamic behavior of the singularities. As reported by some authors [20-22], optical trapping with singularities is a promising branch of physics, and this paper on near-field optics with curved topological features can open new aspects of experimental investigation.

The authors thanks J. M. Hickmann and S. B. Cavalcanti for useful discussions.

\newpage
.

\end{document}